\documentclass[12pt]{article}
\usepackage[latin1]{inputenc}

\usepackage{amsmath}
\usepackage{amsfonts}
\usepackage{amssymb}
\usepackage{graphicx}
\usepackage{geometry}
\usepackage{amssymb,epsfig}
\usepackage{hyperref}

\makeatletter
\renewcommand\section{\@startsection {section}{1}{\z@}%
                                 {-3.5ex \@plus -1ex \@minus -.2ex}%
                                   {2.3ex \@plus.2ex}%
                                   {\normalfont\large\bfseries}}
\renewcommand\subsection{\@startsection{subsection}{2}{\z@}%
                                   {-3.25ex\@plus -1ex \@minus -.2ex}%
                                     {1.5ex \@plus .2ex}%
                                     {\normalfont\bfseries}}
\renewcommand\subsubsection{\@startsection{subsubsection}{3}{\z@}%
                                   {-3.25ex\@plus -1ex \@minus -.2ex}%
                                     {1.5ex \@plus .2ex}%
                                     {\normalfont\itshape}}
\makeatother

\def\pplogo{\vbox{\kern-\headheight\kern -29pt
\halign{##&##\hfil\cr&{\ppnumber}\cr\rule{0pt}{2.5ex}&\ppdate\cr}}}
\makeatletter
\def\ps@firstpage{\ps@empty \def\@oddhead{\hss\pplogo}%
  \let\@evenhead\@oddhead 
}
\def\maketitle{\par
 \begingroup
 \def\thefootnote{\fnsymbol{footnote}}
 \def\@makefnmark{\hbox{$^{\@thefnmark}$\hss}}
 \if@twocolumn
 \twocolumn[\@maketitle]
 \else \newpage
 \global\@topnum\z@ \@maketitle \fi\thispagestyle{firstpage}\@thanks
 \endgroup
 \setcounter{footnote}{0}
 \let\maketitle\relax
 \let\@maketitle\relax
 \gdef\@thanks{}\gdef\@author{}\gdef\@title{}\let\thanks\relax}
\makeatother

\numberwithin{equation}{section}
\newcommand\nn{\nonumber}

\newcommand\eea{\end{eqnarray}}
\newcommand\bea{\begin{eqnarray}}

\newcommand{\p}{\partial}

\def\la{\langle}
\def\ra{\rangle}
\def\beq{\begin{equation}}
\def\eeq{\end{equation}}

\newcommand{\be}{\begin{equation}}
\newcommand{\ee}{\end{equation}}
\newcommand{\bg}{\begin{gather}}
\newcommand{\eg}{\end{gather}}
\newcommand{\bseq}{\begin{subequations}}
\newcommand{\eseq}{\end{subequations}}

\newcommand{\mc}{\mathcal}
\renewcommand{\t}{\tilde}
\newcommand{\pd}[2]{\frac{\partial #1}{\partial #2}}

\newcommand{\cD}{{\cal D}}
\newcommand{\cO}{{\cal O}}
\newcommand{\cS}{{\cal S}}
\newcommand{\cL}{{\cal L}}
\newcommand{\cR}{{\cal R}}
\renewcommand{\l}{L}
\newcommand{\PIR}{\Psi_{\rm IR}}
\newcommand{\PUV}{\Psi_{\rm UV}}

\newcommand{\WUV}{W_{\rm UV}}
\newcommand{\hWUV}{\hat W_{\rm UV}}
\newcommand{\WIR}{W_{\rm IR}}
\newcommand{\phiUV}{\phi_{\rm UV}}
\newcommand{\aUV}{a_{\rm UV}}
\newcommand{\LUV}{L_{\rm UV}}
\newcommand{\tphi}{\tilde\phi}
\newcommand{\ba}{\bar a}
\newcommand{\ta}{\tilde a}

\newcommand{\tpi}{\tilde\pi}
\newcommand{\hg}{\hat g}
\newcommand{\shg}{\sqrt{\hat g}}
\newcommand{\hR}{\hat{\cal R}}

\begin{document}

\setcounter{page}0
\def\ppnumber{\vbox{\baselineskip14pt
}}
\def\ppdate{\footnotesize{SLAC-PUB-15249 ~~~ SU-ITP-12/24}} \date{}

\author{Xi Dong,$^{a,b}$ Bart Horn,$^{a,b,c}$ Eva Silverstein,$^{a,b}$ Gonzalo Torroba$^{a,b}$\\
[7mm]
{\normalsize \it $^a$Stanford Institute for Theoretical Physics }\\
{\normalsize  \it Department of Physics, Stanford University}\\
{\normalsize \it Stanford, CA 94305, USA}\\
[7mm]
{\normalsize \it $^b$Theory Group, SLAC National Accelerator Laboratory}\\
{\normalsize \it Menlo Park, CA 94025, USA}\\
[7mm]
{\normalsize \it $^c$Physics Department and Institute for Strings, Cosmology and Astroparticle Physics}\\
{\normalsize \it Columbia University, New York, NY 10027, USA}
}

\bigskip
\title{\bf Moduli Stabilization and the Holographic RG for AdS and dS 
\vskip 0.5cm}
\maketitle
%

\begin{abstract}

We relate moduli stabilization ($V'=0$) in the bulk of $AdS_D$ or $dS_D$ to basic properties of the Wilsonian effective action in the holographic dual theory on $dS_{D-1}$: the single-trace terms in the action have vanishing beta functions, and higher-trace couplings are determined purely from lower-trace ones.  In the de Sitter case, this encodes the maximal symmetry of the bulk spacetime in a quantity which is accessible within an observer patch. Along the way, we clarify the role of counterterms, constraints, and operator redundancy in the Wilsonian holographic RG prescription, reproducing the expected behavior of the trace of the stress-energy tensor in the dual for both $AdS_D$ and $dS_D$.   We further show that metastability of the gravity-side potential energy corresponds to a nonperturbatively small imaginary contribution to the Wilsonian action of pure de Sitter, a result consistent with the need for additional degrees of freedom in the holographic description of its ultimate decay.      

\end{abstract}
\bigskip
\newpage

\tableofcontents

\vskip 1cm

\section{Introduction}\label{sec:intro}

The extensive observational evidence for inflation and dark energy strongly motivates the development of a more complete framework for de Sitter spacetime and its decays.  Perhaps the most conservative approach to the problem is to formulate the physics inside an observer patch as in~\cite{dSdS,dSdSbrane,FRW,AHM}.  In addition to restricting attention to operationally measurable quantities, this builds the holographic dual from a unitary, Lorentzian low energy theory whose count of degrees of freedom provides an estimate of the de Sitter entropy.   On the other hand, this approach does not make manifest the symmetries of global de Sitter spacetime, a feature which in the dS/CFT approach to the problem~\cite{dSCFT} immediately implies conformal invariance of the dual.  However, the symmetries follow from the stabilization of the scalar moduli (leading to a maximally symmetric solution), and as we will see, the consequences of this  {\it are} evident within a causal patch.

In general, one would like to understand what specifications are required of the matter sector\footnote{We say {\it matter sector} here because in both approaches, dynamical lower-dimensional gravity (integration over metrics) is required to complete the calculation of observables, a complication intrinsic to the physics of de Sitter space.  This complication is minimal in low dimensional examples, and in all dimensions ultimately disappears in the nonperturbative decay of metastable de Sitter~\cite{FRW,HarlowSusskind}.  Despite the ultimate decay of de Sitter space (at least in known UV complete constructions), it is still worth understanding as much as possible about the very long-lived de Sitter phase itself.                      
} in the dual in order for it to reconstruct the physics of an observer patch of dS.\footnote{A brane construction which addresses this implicitly in a particular example is given in~\cite{dSdSbrane}.}  Because of its large number of degrees of freedom, the matter sector induces the $D-1$ dimensional Planck mass, which is parametrically far above the scales that we focus on in this work.  Our main goal is to determine concrete properties of the matter sector that is coupled to this residual gravity.  Using the framework of the holographic Wilsonian renormalization group developed in~\cite{HP},\footnote{following many previous works~\cite{Mansfield:1999kk,de Boer:1999xf,Li:2000ec}. See also~\cite{Akhmedov:2010sw,Faulkner:2010jy} for more recent related work.} we establish a simple and general feature of the dual in the dS/dS correspondence \cite{dSdS,dSdSbrane}:\  its single-trace couplings have vanishing $\beta$ functions.  Multiple trace terms in the Wilsonian action do run but in a special way dictated by the maximal symmetry of the bulk spacetime.  Moreover, the holographic RG reproduces the expected behavior of the trace of the stress-energy tensor as we will discuss further below.  

It will be very interesting to apply this lesson to help construct and elucidate specific dual field theories, including the concrete examples in~\cite{dSdSbrane}.  
The vanishing of $\beta_{\text{single-trace}}$ holds for each direction in scalar field space which is metastabilized; this will allow us to analyze it in simpler constructions which uplift the AdS/CFT potential without metastabilizing all the moduli \cite{inprogress}.                        

In the fully metastabilized case, we will work in the dS/dS framework~\cite{dSdS}.  This follows at a macroscopic level from the metric
\beq\label{dSdSmet}
ds^2_{dS_D} = dy^2+\sin^2\frac{y}{R} ds^2_{dS_{D-1}}
\eeq
exhibiting $dS_D$ as a warped compactification down to $dS_{D-1}$ with two highly redshifted regions indicating two low energy theories coupled to each other and to $D-1$ dimensional gravity.  As explained in~\cite{dSdSbrane}, this same structure arises in a simple way more microscopically when one uplifts Freund-Rubin AdS/CFT solutions to de Sitter using contributions to the moduli potential which arise in string theory.  The nontrivial agreement between the dS/dS metric (\ref{dSdSmet})  and the basic structure of the uplifted brane construction is encouraging, motivating further development of the dual descriptions.      

In using the framework of the holographic Wilsonian RG, we path integrate first over the fields in the bulk (separating them into UV and IR pieces), leaving for last the integration over the fields -- including gravity -- at the UV slice (the central slice between the two throats).  This last path integral includes integration over the $D-1$ dimensional metric as well as other sources which generically induce couplings between the two throats.  The RG properties of the matter sector that we will determine are what they would be for a stand-alone matter theory (without couplings to gravity or to the other throat), because we have left the path integral on the UV slice for the last step.  However, these RG properties are the most useful in the sense that they could guide us in building putative dual theories sector-by-sector before coupling them together and to gravity.  The situation is similar to studying the Standard Model beta functions in particle physics and neglecting contributions from gravity and hidden sectors, assuming that these couplings are weak.

Although the full de Sitter symmetries are not as manifest within the static patch, they follow from
a simple feature which {\it is} evident in the region accessible to a single observer:\ the stabilization of the scalar fields $\phi_I$ in the system.  In AdS/CFT, this translates into vanishing $\beta$ functions for the couplings dual to the scalar fields $\phi_I$ as well as for all operators generated in the RG flow \cite{HP}; conversely, solutions with radially rolling scalars describe nontrivial RG flows in the dual field theory.
In this work we will generalize this to the de Sitter observer patch, using the holographic renormalization group to translate the existence of
a stable or metastable minimum of the potential $V(\phi_I)$ to basic simplifications of the dual theory.  We find that the single-trace couplings in the holographic dual do not run, and that the higher-trace couplings are determined purely from lower-trace ones in the Wilsonian action.  

Finally, we will analyze the (likely general) case that the de Sitter phase is only meta-stable, building in a runaway direction in $V(\phi_I)$ as occurs in string-theoretic de Sitter constructions.  This introduces bounce solutions into the semiclassical calculation of the Wilson action, leading to an exponentially suppressed imaginary part.  This is consistent with the expectation that pure dS is not a complete theory in itself~\cite{HarlowSusskind}; indeed, additional degrees of freedom come into play in formulating its decays~\cite{FRW, FRWCFT}.  

This gives a new application of the holographic RG, which yields results that were not known in any other way.  As emphasized in~\cite{HP}, although it roughly corresponds to a Wilsonian prescription of integrating out high energy modes, the precise implementation of this scheme in the traditional field theory variables is not understood; it is a kind of functional RG~\cite{functionalRG} but not precisely the same as those formulated in field theory.  In any case, it will be very interesting to turn things around and analyze in field theory what is required to obtain the structures derived here (the vanishing of the single trace $\beta$ functions and the specific form of the running of mulitrace terms) from the field content of candidate de Sitter duals, including~\cite{dSdSbrane}.

This paper is organized as follows.  In \S \ref{sec:results}, we derive in a simple way two of our basic results, that single-trace $\beta$ functions vanish in the Wilsonian holographic RG and that the effective action for higher-trace terms is determined by lower-trace terms.  In  \S \ref{sec:3d}, we elucidate the redundancy of the trace of the stress-energy tensor in this framework.  For the reader interested in the main results, these two sections are sufficient. In \S \ref{sec:framework}, we lay out the general Hamilton-Jacobi framework 
for holographic RG, applying~\cite{HP} to more general foliations, in particular the $dS_{D-1}$ case, and to arbitrary zero or nonzero modes of the fields.  Then in \S \ref{sec:appl} we derive the consequences of moduli stabilization for the Wilsonian action for scalar operators and the trace of the stress tensor.  In \S \ref{sec:meta} we briefly discuss the consequences of the metastability of the potential for the holographic RG, and we close in \S \ref{sec:concl} with some further comments.  In the appendices we work out some details on the stress-energy tensor in the Wilsonian action, as well as an explicit calculation in $D=3$.

\section{Basic results:  path integral derivation}\label{sec:results}

Let us start by explaining our basic framework and results.
Much of this is in direct parallel with the analysis in~\cite{HP}, generalized appropriately to the case of a $dS_{D-1}$ foliation of the bulk.  Along the way, we will need to clarify the role of the warp factor and include appropriate counterterms to obtain standard AdS/CFT operator redundancies, applying \cite{Trefs, de Boer:1999xf}\ to the Wilsonian Holographic RG.  Our main result is that the scale-invariance of the single-trace couplings is a consequence of moduli stabilization for dS as well as AdS, and that in both cases the higher-trace couplings are determined in terms of the lower-trace ones.      

\subsection{Framework}

Since we are interested in the consequences of moduli stabilization, we will focus on the dynamics of a bulk scalar field $\phi$ dual to an operator ${\cal O}$ in the $d\equiv D-1$ dimensional dual field theory, along with its effect on the warp factor $a(y,x)$ in the gauge-fixed metric
\beq\label{metric}
ds^2=dy^2 + a(y,x)^2 \hg_{\mu\nu} dx^\mu dx^\nu . \,
\eeq
This system has Euclidean-signature action
\beq\label{Sscalar}
\cS= \int dy\, d^dx \, a^d \shg \left\{ \frac{1}{2} \left(\pd{\phi}{y}\right)^2 + \frac{1}{2a^2} \hg^{\mu\nu} \p_\mu\phi \p_\nu\phi + V(\phi) -\frac{1}{2}(\cR+ \mc L_{GHY}) + \frac{1}{a^d}\partial_y(a^d\cL_{CT}) \right\} \,,
\eeq
where $\cL_{CT}$ is a counterterm Lagrangian, a local function of $\phi$, the metric, and their derivatives in the $x$ directions.\footnote{We have chosen coordinates \eqref{metric}\ here; the counterterm could be written more generally as $n^\mu\partial_\mu(a^d \cL_{CT})$. The factor of $a^d$ in $\cL_{CT}$ is added to simplify our formulas below.}  For the case of AdS/CFT, the counterterms were derived in \cite{Trefs}.  For the dS case, we will fix $\cL_{CT}=0$ using the symmetries in \S\ref{sec:3d}. The bulk scalar curvature $\cR$ and the Gibbons-Hawking-York term $\mc L_{GHY}$\footnote{Here we have chosen to write the Gibbons-Hawking-York boundary term as a total derivative in the bulk action.} combine to give
\be\label{Rprime}
\cR+ \mc L_{GHY} = d(d-1) \frac{1}{a^2} \left(\pd{a}{y}\right)^2+(d-1)(d-2) \frac{1}{a^4}\hat g^{\mu\nu} \partial_\mu a \partial_\nu a+\frac{1}{a^2}\hR
\ee
where $\hR$ is the Ricci scalar of the $d$-dimensional metric $\hg_{\mu\nu}(x)$.  
One can similarly include the dynamics of transverse traceless modes of the graviton, but we will focus on the scalar fields and their interaction with the warp factor $a(y,x)$.   
The warp factor itself is determined by a constraint equation -- it is not an independent dynamical degree of freedom on the gravity side.  This corresponds to the redundancy of the trace of the stress-energy tensor in the dual field theory, a feature we will recover in our framework.     

In the holographic Wilsonian RG as formulated in~\cite{HP}, one uses the scale-radius duality
\beq\label{scalerad}
E = E_{\text{proper}}a(y, x)
\eeq
to map the integration over high energy modes on the field theory side to integration over the fields at large warp factor $a$, which will correspond in our coordinates to large radial position $y$ on the gravity side.  For supergravity modes the proper energy scales like $1/R$ in terms of the bulk curvature radius $R$, for strings the proper energy is of order the square root of the string tension, and additional scales may arise in general.\footnote{One of the remaining subtleties with holographic RG is the fact that the gravity side contains excitations with different proper energies, so a cutoff at $y=L$ is not in fact a cutoff on energy scales in the dual QFT.  This may be a feature rather than a bug, potentially suggesting a novel way to organize the path integral in QFT.}  To capture the physics of the lightest scalar `moduli' fields, the supergravity scale will be most relevant.     

Let $y=\LUV$ be the most UV slice
in our geometry (i.e.\ the slice with the largest warp factor $a$).\footnote{In the AdS case, we can regulate this as in \cite{HP}.  In the dS case, we first consider a single warped throat, say the one with $0\le y\le \pi R/2$ in (\ref{dSdSmet}).  In \S\ref{sec:3d}, we will discuss its coupling to the full causal patch.}  To formulate our renormalized theory, we introduce an arbitrary scale   by choosing 
an intermediate radial position $y=L$ with respect to which we will divide the path integral into high and low energy degrees of freedom.  Defining $\bar a(y)$ to be the classical warp factor in AdS or dS (or more general geometries), this arbitrary energy scale will be of order $\mu_L\sim \bar a(L)/R$.
More precisely, we work at fixed proper distance $\LUV-L$ from the UV slice of our geometry.  The gravity path integral is divided into a UV part with fields integrated over $L< y <\LUV$, and IR piece from the region $0< y<L$, and an integral over the fields at the surface $y=L$:
\bea\label{eq:Zgravity}
Z &=& \int \cD\t a\cD \t \phi\,\int \cD a\cD \phi|_{y>L} \exp(-\kappa^{-2}\cS|_{y>L}) \,\int \cD a \cD \phi|_{y<L} \exp(-\kappa^{-2}\cS|_{y<L}),
\eea
where $\phi(x,L) = \t \phi(x)$ and $a(x,L)=\tilde a (x)$. Here $\kappa^2 \sim G_N$, with $\kappa^2 \to 0$ corresponding to the planar limit in the holographic dual.  We will work in this semiclassical approximation.

The UV part of the path integral
\beq\label{PsiUV}
\PUV(\tphi,\t a,\l)= \int \cD a \cD \phi |_{y>\l} \exp(-\kappa^{-2}\cS|_{y>\l}) \,,\quad
\eeq
is evaluated with radial boundary conditions $\phi(x, \LUV) = \phiUV(x)$, $a(x, \LUV) = \aUV(x)$ and $\phi(x,L)=\tilde\phi (x)$, $a(x,L)=\tilde a(x)$.  $\PUV$ can be constructed equivalently via radial Hamiltonian evolution from the boundary, as we will describe in more detail in later sections.  
Let us use $\cS^{(0)}$ to denote the bulk action \eqref{Sscalar} with every term except the counterterm.  We have
\begin{align}\label{PsiUV2}
\PUV(\tphi,\ta,\l) =& \int \cD a \cD \phi |_{y>\l} \exp\left\{-\kappa^{-2} \left(\cS^{(0)}|_{y>\l}+\cS_{CT}[\phiUV,\aUV]-\cS_{CT}[\t\phi,\t a] \right)\right\} \\
=& e^{\kappa^{-2} \cS_{CT}[\t\phi,\t a]} \PUV^{(0)} e^{-\kappa^{-2} \cS_{CT}[\phiUV,\aUV]} \,.\quad
\end{align}

The IR part of the path integral
\beq\label{PsiIR}
 \PIR(\tphi,\tilde a,\l)= \int \cD a \cD  \phi|_{y<\l} \exp(-\kappa^{-2}\cS|_{y<\l}) = 
 \PIR^{(0)} e^{-\kappa^{-2} \cS_{CT}[\t\phi,\t a]}  
\eeq   
(with boundary conditions $\phi(x,L)=\tilde\phi (x)$, $a(x,L)=\tilde a (x)$)
is postulated \cite{HP}\ to be of the form
\beq\label{PsiIRQFT}
\PIR=\int \cD M|_{E<\tilde a/R}\,e^{-S_0[M,\hg] +\kappa^{-2} \int d^dx \, \ta^d \shg \, \tphi \cO}\,,
\eeq
in terms of the microscopic fields $M$ of the holographic dual.  In this formulation the scale $\tilde a/R$ (which is semiclassically $\mu_L\sim \ba(L)/R $) is some form of UV cutoff on the fields that we integrate over to construct this low-energy part of the path integral.  As in \cite{HP}, we will not solve the problem of making this explicit in the microscopic field theory variables, but will use the gravity side to investigate the structure of the RG and its relation to moduli stabilization given the postulate (\ref{PsiIRQFT}).
According to this conjecture, which follows naturally from the UV/IR relation in AdS/CFT, $\PIR$ is a cutoff version of the field theory partition function.  
As $L\to \LUV$, $\PIR$ approaches the field theory partition function with sources $\phiUV$, $\aUV$, and $\PUV$ becomes a delta function localized on $\phiUV$, $\aUV$ \cite{HS}.

In \eqref{PsiIRQFT} $\mc O$ is a single-trace operator dual to $\phi$.  Correlators of the trace $T$ of the stress-energy tensor are obtained by differentiating with respect to $\log(\ta)$.  We will see explicitly below that in Poincar\'e AdS/CFT, our prescription yields $T=0$, and that we also obtain the correct results for correlation functions of $T$ for AdS and dS with de Sitter slicing.  

The role of the warp factor -- ultimately a non-dynamical redundant variable -- is somewhat complicated in the expression \eqref{PsiIRQFT}.  Note that we have chosen to put the fluctuation of the trace part of $d$-dimensional metric into $\ta$ instead of $\hg_{\mu\nu}$.
Equivalently, we may write \eqref{PsiIRQFT} in a more standard form by putting the fluctuation of $\ta$ away from its background value $\ba(L)$ into the trace part of the metric $\hg_{\mu\nu}$ on which the field theory lives, via the following change of variables:
\beq\label{gprime}
\hat g'_{\mu\nu}=\frac{\ta^2}{\ba(L)^2} \hat g_{\mu\nu} \,,
\eeq
therefore removing the fluctuation of $\ta$ from the cutoff on microscopic fields.
Here again $\bar a(y)$ is the background warp factor in AdS or dS (or more general geometries with radial evolving scalars in the case that we do not stabilize the moduli).
This expresses $\PIR$ more clearly as a cutoff version of the QFT partition function:
\begin{align}\label{PsiIRQFTprime} 
\PIR=&\int \cD M|_{E<\mu_L}\,e^{-S_0[M,\hg']+\kappa^{-2} \int d^dx \, \ba(L)^d \sqrt{\hg'}\tphi \cO}\\
=&\int \cD M|_{E<\mu_L}\,e^{-S_0[M,\hg] +\kappa^{-2} \int d^dx \, \ba(L)^d \sqrt{\hg'} \,\tphi \cO +\kappa^{-2} \int d^dx \, \ba(L)^d \sqrt{\hg} \,\left(\frac12 \delta\hg'_{\mu\nu}T^{\mu\nu}+\ldots\right)}\,,
\end{align}
where again the cutoff is $\mu_L\sim \ba(L)/R$, and on the second line we have simply expanded $S_0[M,\hg']$ to linear order around the metric $\hg$ which has a non-fluctuating trace part, and the terms in `$\ldots$' refer to the nonlinear couplings of the stress-energy tensor and the metric fluctuation $\delta\hg'_{\mu\nu}=\hg'_{\mu\nu}-\hg_{\mu\nu}$.   
Focusing on the fluctuation of the warp factor (i.e.\ the trace part of the metric fluctuation\footnote{The traceless fluctuations of the metric can be treated in analogy with scalars $\phi$.}), we may further rewrite \eqref{PsiIRQFTprime} as
\beq\label{PIRQFT}
\PIR=\int \cD M|_{E<\mu_L}\,e^{-S_0[M,\hg] +\kappa^{-2} \int d^dx \sqrt{\hg} \left[\ta^d \tphi \cO +\ba(L)^{d-1} \delta\ta \, T+\ldots\right]}\,,
\eeq
where we have used \eqref{gprime} again.

Now we can substitute \eqref{PIRQFT} into \eqref{eq:Zgravity} and integrate over $\tilde a$ and $\tilde\phi$, postponing the path integral over the microscopic fields $M$.  This gives a prescription \cite{HP}\ for a holographic Wilsonian effective action $s({\cal O}, L)$
\beq\label{WilsonI}
\exp\left(-\kappa^{-2} s(\cO,\l)\right)= \int \cD \t a \cD \tphi\, \PUV(\tphi,\t a, \l) \, e^{\kappa^{-2} \int d^dx \sqrt{\hg} \left[\ta^d \tphi \cO +\ba(L)^{d-1} \delta\ta \, T+\ldots\right]} \,.
\eeq
This action is guaranteed to be local on energy scales much lower than $\mu_L$; at the semiclassical level it is the Legendre transform of $\PUV$ which can be constructed by radial evolution from the boundary via a local Hamiltonian (as will be worked out in detail in \S\ref{sec:framework}).  It is interesting to note that the Legendre transform of the Wilsonian effective action -- which is analogous to $\WUV\equiv \kappa^2\log\PUV$ -- also plays a role in studies of the exact RG formalism in standard quantum field theory \cite{functionalRG}.  

\subsection{Consequences of moduli stabilization}

We will be concerned with special properties possessed by the holographic Wilsonian action \eqref{WilsonI} when the bulk scalar field has a potential with a stable 
or metastable extremum at $\phi=\phi_*$ of the form
\beq\label{Vexp}
V(\phi)=V _*+\sum_{n=2}^{\infty} \frac{1}{n!} V^{(n)}_* (\phi-\phi_*)^n \,.
\eeq
To see the main effect of the local minimum in $V(\phi)$, let us focus on zero modes of the fields, taking them independent of the $x$ directions (the directions along the $d$-dimensional slice).  Our conclusions will not depend on this, as we will show by using a more detailed systematic analysis in \S \ref{sec:appl}.  As already mentioned, we will work semiclassically, taking $\kappa^2 \to 0$, so that the path integral (\ref{PsiUV}) is dominated by saddle points. 

Let us choose the UV boundary condition $\phi(\LUV)=\phi_*$ and $a(\LUV)=\bar a(\LUV)$.   As a warmup, consider first the scalar field on a fixed background warp factor $\ba(y)$. The path integral for $\PUV$ will be dominated by the classical solution that describes the field rolling on the inverted potential from $\tilde\phi$ to $\phi_*$ (with ``Hubble'' friction coming from the $y$-derivative of the warp factor).  This classical solution can be expanded in $\tilde\phi-\phi_*$:
\beq\label{phiexp}
\phi(y)= \phi_*+\phi_1(y)(\tilde\phi-\phi_*)+\phi_2(y)(\tilde\phi-\phi_*)^2+\dots 
\eeq 
with the first term encoding the fact that the solution becomes the constant $\phi(y)=\tilde\phi=\phi_*$ as the two boundary conditions $\tilde\phi$ and $\phi_*$ are brought together. 
When we plug this solution back into the bulk action \eqref{Sscalar}, each term depends on the fluctuation $\phi-\phi_*$ only via quadratic terms and higher.\footnote{The counterterm Lagrangian $\cL_{CT}$ can be chosen so that it does not have a linear term in $\tilde\phi-\phi_*$.}  This implies the following form for the semiclassical UV path integral:
\beq\label{WUV}
\PUV(\tphi,\l) = \exp\left\{-\kappa^{-2}  \int d^dx \, \ba^d(\l) \shg \, \left[ w_0(L)+\frac{1}{2}w_2(L) (\tilde\phi-\phi_*)^2+\dots\right]\right\} \,.
\eeq
In particular, there will be no linear term in $\tphi-\phi_*$ (i.e.\ $w_1=0$).  In \S \ref{sec:framework}, we will recover the same result in a systematic analysis by expanding the Hamilton-Jacobi equation for the radial evolution of $\PUV$.  

To obtain the holographic Wilsonian action, we plug $\PUV$ into the integral transform \eqref{WilsonI}.  The saddle point solution is $\tilde\phi-\phi_*=\frac{\cO}{w_2}+\dots$, where `$\dots$' denotes higher powers of $\cO$.  This leads to a Wilsonian action
\beq\label{WilsI}
s(\cO,\l)=\int d^d x \, \ba(L)^d \sqrt{\hat{g}}\left(w_0(\l) - \phi_* \cO - \frac{1}{2 w_2(\l)}\cO^2 +\dots\right) \,,
\eeq  
where `$\dots$' includes higher-order terms in $\cO$.  In general, we can define a set of couplings $\sigma_n(\l)$ by expanding the Wilsonian action in powers of $\cO$
\beq\label{Wilexp}
s(\cO,\l)=\int d^d x \, \ba(L)^d \sqrt{\hat{g}} \sum_{n=0}^{\infty} \frac{1}{n!} \sigma_n(\l) \cO^n \,
\eeq
where $\sigma_n(\l)$ can be read off from \eqref{WilsI}.  In particular, we note that $\sigma_1=-\phi_*$ is independent of $\l$.

As a final step, let us rewrite this in standard field theory conventions, using
\beq\label{QFTrelns}
\mu_L=\frac{\ba(L)}{R} \,,\qquad
\cO_{QFT}=\frac{\ba(L)^\Delta}{R^{\Delta-(d+1)/2}} \frac{\cO}{\kappa} \,,\qquad
T_{QFT}=\ba(\l)^d \frac{T}{\kappa^2}
\eeq
to obtain a Wilsonian action of the form    
\beq\label{WilsIrescaled}
\kappa^{-2} s(\cO_{QFT},\mu_L) = \int d^d x \shg \sum_{n=0}^{\infty} \left(\frac{R^{d-1}}{\kappa^2}\right)^{1-n/2} \frac{1}{n!} \hat\sigma_n(\l) \mu_L^{d-n\Delta}\cO_{QFT}^n \,,
\eeq  
where we have defined the dimensionless couplings $\hat\sigma_n=R^{1-n}\sigma_n$ following the usual convention in quantum field theory.  Their running as a function of $\mu_\l$ defines beta functions.
Here the dimensions $\Delta$ are obtained from the scalar masses using the standard AdS/CFT dictionary; this applies to the de Sitter case as well by working in the low-energy region where dS/dS reduces to AdS/dS \cite{dSdS}.\footnote{Here we relate masses $m$ to dimensions $\Delta$ through the curvature radius, which we take to be the same for dS and the fiducial AdS theory which reduces to it at low energies.}   

If we consider multiple scalars, this result continues to hold in each direction in field space with
a local extremum; since the potential starts out quadratic there is no flow in the single-trace coupling
for the corresponding operator $\cal O$.  This will be useful in connecting our results to UV complete uplifts of AdS/CFT \cite{inprogress}, since it is significantly simpler to obtain partial moduli stabilization than to work with a complete de Sitter construction.    

From (\ref{WilsI} -- \ref{WilsIrescaled}) we see that the single-trace coupling $\sigma_1$ does not run: it is fixed at $-\phi_*$ and does not depend on $L$ (or equivalently $\mu_L$).   Furthermore, the coefficient $\sigma_n$ of the $\cO^n$ term in the Wilsonian action \eqref{Wilexp} is determined by $w_m$'s with $m\le n$; this can be seen by working out the saddle point solution for $\tphi-\phi_*$ order by order in $\cO$.  The $w_n$ can themselves be determined order by order from the saddle point solution.  Therefore, a given coupling $\sigma_n$ can be determined as a function of $\l$ once we know the lower ones $\sigma_{m<n}$, indicating an iterative structure in the RG flow.

Both of the above results are substantial simplifications of the RG evolution.  In standard functional RG in quantum field theory~\cite{functionalRG}, the flow of a generic coupling gets contributions from both lower and higher dimension operators, and in matrix theories from both higher- and lower-trace terms.

These results continue to hold when we include backreaction on the scale factor $a(y,x)$.  In this case, the coefficients $\phi_i$ in the expansion \eqref{phiexp} develop dependence on $\tilde a$, but they still have the property noted above that the solution is $\phi(y)=\phi_*$ if we take the boundary condition $\tilde\phi$ to $\phi_*$.  In addition, the classical solution for $a(y)$ can also be expanded in powers of $\tphi-\phi_*$:
\beq\label{aexp}
a(y)=a_0(y)+a_2(y)(\tphi-\phi_*)^2+\dots,
\eeq
where we have left implicit the dependence on the boundary condition $\ta$.  We note that there is no linear term in \eqref{aexp}.  This is because the equation of motion for $a(y)$ (i.e.\ the second-order Friedmann equation in Euclidean signature) is sourced by $V(\phi)+\frac{1}{2}\phi'(y)^2$, which does not have a linear term in $\tphi-\phi_*$ when we use \eqref{phiexp}. 
As a result, the bulk action \eqref{Sscalar} does not have any term linear in $\tphi-\phi_*$, and the UV path integral (\ref{PsiUV}) takes the form
\beq\label{WUVa}
\PUV(\ta,\tphi,\l) =\exp\left\{-\frac{1}{\kappa^2}  \int d^dx \, \ta^d \shg \left[ w_0(L,\ta)+\frac{1}{2}w_2(L,\ta) (\tilde\phi-\phi_*)^2+\dots\right]\right\} \,,
\eeq
where the coefficient functions $w_n$ can be further expanded as
\beq\label{wexpI}
w_n(L,\ta)=w_{n0}(L)+w_{n1}(L)(\ta-\ba(L))+\dots  
\eeq
with a generally nonzero linear term in $\ta-\ba(L)$.  As we will show in Appendix \ref{app:T0}, this linear term does not produce single-trace couplings for $\cO$.  Therefore, we retain the feature that the single-trace action for $\cO$ is uncorrected under the holographic RG flow.  In our systematic analysis in \S\S \ref{sec:framework} and \ref{sec:appl}, we will set up the equations determining the multi-trace couplings, and note some further simplifications in their structure which result from the maximal symmetry of the warp factor.

Nothing about this derivation depended on the sign of $V_*$ or the shape of the holographic screen, applying for arbitrary slicing $a(y)$.
Our main result is indeed that the dS/dS dual theory living on $dS_{D-1}$ must have special field content and interactions which guarantee the cancellation of the $\beta$ functions for  single-trace couplings.  At the same time, multi-trace couplings do flow in a particular way, with the iterative structure mentioned above.  It will be very interesting to return to the brane construction  of \cite{dSdSbrane}\ to understand how these special features come about there.  More generally, one may be able to use this structure to construct new theories with the right properties to provide duals for the de Sitter static patch.

Such dual theories, however, are likely to be ultimately incomplete because of non-perturbative decays of de Sitter space.
In \S \ref{sec:meta} we will see this in the holographic RG itself due to additional solutions (related to `bounce' solutions in instanton physics) which contribute to the path integral and lead to imaginary contributions to the Wilsonian action.  This is in accord with the arguments \cite{HarlowSusskind, FRW} that a complete description of de Sitter space must ultimately include its decays.  There is a precedent for such behaviors already in the study of AdS/CFT and warped compactifications:\ freely-acting supersymmetry-breaking orbifold theories \cite{orbifold}\  are only perturbatively stable.  When we impose a cut-off by embedding them in a warped compactification, these theories are metastable; they survive for exponentially long (as a function of large inverse couplings) but ultimately decay.  Although not completely self-contained, their field content and couplings are known, given by specific quiver gauge theories.          
It would be interesting to use the results of the present paper to constrain the specific field content and couplings of the analogous theories dual to the de Sitter static patch.

\section{Operator relations and counterterms}\label{sec:3d}

Let us analyze further the role of the warp factor $a(y,x)$ and its dual operator $T$, the trace of the stress-energy tensor,  exhibiting their redundancy in the holographic Wilsonian RG.  This will clarify the general framework, reproducing the trace anomaly for $AdS_D$ and leading to a simple result for the $dS_D$ case.  

From the construction introduced above, correlation functions of $T$ (and $\cO$ when it is nontrivial) are given by (using \eqref{PIRQFT})
\beq\label{Tcorr}
\langle T^n \cO^m \rangle = \frac{1}{Z} \int \cD\tilde a \cD\tilde\phi\, \PUV\left(\frac{1}{\ba(\l)^{d-1}\shg}\frac{\delta}{\delta\ta}\right)^n\left(\frac{1}{\t a(\l)^d\shg}\frac{\delta}{\delta\t\phi}\right)^m\PIR \,.
\eeq
One may analyze this directly in terms of $\PUV$, related to the Wilsonian action by a Legendre transform, or in terms of the Wilsonian action itself.   
In writing \eqref{Tcorr}, we must keep in mind that only low-energy modes of the operators $T$ and $\cO$ are  
to be inserted, as specified in \eqref{PIRQFT}.  In particular, it is important to study low-energy correlators which can be reliably computed after having done the path integration over UV modes which led to $\PUV$.    

We could use the relation between bulk and boundary observables derived in \cite{bulkobs}\ to construct low-lying operators very precisely.   The smeared operators 
defined in \cite{bulkobs}\ correspond to bulk fields; these are QFT operators $\cO$ convolved with their corresponding bulk-to-boundary propagator to give $\phi(x,y)= \int dy d^dx' K(x-x',y)\cO(x')$.
If we insert these fields into the path integral in the IR region $y<L$ of our geometry, 
they do not affect the path integration over the UV region $y>L$ at all.  Therefore, correlators of the smeared operators provide explicit low-energy observables which are precisely calculable after integrating out the UV modes in the holographic Wilsonian RG.  Analyzing these correlators in detail would require folding in the appropriate bulk-to-boundary propagators, including mixing between scalars and the warp factor.  Because the Wilsonian prescription defined above yields exactly the same path integral calculating these observables as one gets using the usual AdS/CFT dictionary (since the insertions do not interfere with the Wilsonian order of integration), this calculation will automatically precisely yield the correct trace anomaly.      

More generally, even if we do not define the low-lying modes precisely in this way, correlators of low-energy modes of the operators must satisfy the correct operator identities up to corrections arising as a power series in $(\text{energy}/\mu_L)$.  We can see this directly from our prescription in \S\ref{sec:results}.  With our counterterm prescription in place, 
$\PIR$ is the partition function for the QFT cut off at the scale $\mu_L$  (and $\PUV$ approaches $\delta(\ta-\aUV)\delta(\tphi-\phiUV)$ as $\mu_L\to\infty$).  $\PIR$ necessarily generates the appropriate operator relations as we take $\mu_L\to\infty$ relative to the energy scale at which we work, since in that limit we recover the standard AdS/CFT partition function (the latter was analyzed explicitly in \cite{Trefs}).  

\subsection{$(A)dS_3/X_2$ case}

It is interesting to see how this works explicitly in the simplest cases.  
Let us work it out for $d=2$, starting with a fixed scalar background $\phi=\phi_*$, where $\phi_*$ is an extremum of the bulk potential $V(\phi)$.  
In $d=2$, the bulk action \eqref{Sscalar} restricted to the $d$-dimensional zeromodes becomes quadratic in the scale factor $a$, and we can easily obtain the UV and IR amplitudes by explicitly evaluating the path integrals \eqref{PsiUV}\ and \eqref{PsiIR}, which are Gaussian for this case.  We will then determine the required counterterm Lagrangian, following \cite{Trefs}\ for the $AdS_D$ case and using symmetries in the $dS_D$ case.  
From this, we can determine the correlation functions of $T$ at scales below $\mu_L$ and check that $T$ is redundant.   

 In $d=2$, the bulk action \eqref{Sscalar}\ without counterterms at fixed $\phi=\phi_*$ becomes
\beq\label{Sa2d}
\cS^{(0)}= \int dy \int d^2x \, \shg \left[ -\left(\pd{a}{y}\right)^2 +a^2 V_* -\frac{1}{2}\hR \right] \,,
\eeq
with the $y$ integral ranging from 0 (or $-\infty$) to $L$ for $\PIR$, and from $L$ to $\LUV$ for $\PUV$.  
We would first like to compute the path integral \eqref{PsiIR}\ for $\PIR$ with the boundary condition $a(\l)=\ta$.  We will focus on the zero mode, assuming $\ta$ to be independent of $x$, but higher harmonics in the $x$ directions work via similar Gaussian integration.\footnote{Decomposing these into eigenmodes of the $d$-dimensional Laplacian shows that their effect can be packaged as additional terms in $a^2 V(\phi = \phi_*)$ which are extremized when the higher harmonics vanish.} 
Since the bulk action $\cS$ is quadratic in $a$ with a wrong-sign kinetic term, we implicitly use the prescription that rotates the integration contour $a\to ia$ on the complex plane, so that the path integral becomes convergent.  The integral is then given by the action evaluated at the unique classical trajectory (satisfying the equations of motion including the Hamiltonian constraint with the boundary condition just noted).

Let us begin by considering the simplest case of Poincar\'e AdS/CFT, for which we wish to recover the standard result that the trace of the stress-energy tensor vanishes:  we should find that $T=0$ in correlation functions at least for low-energy modes of $T$.  In this case the appropriate classical solution is
\beq\label{aPoinc}
a(y)=\tilde a e^{(y-L)/R},
\eeq
where the AdS radius $R$ is related to $V_*$ by $V_*=-d(d-1)/2R^2$.
From this one finds (in the limit $\kappa^2 \rightarrow 0$ \footnote{As long as we are working in this limit we can ignore the prefactor from the functional determinant, which is independent of $\t a$.  The path integral can then be evaluated in arbitrary dimension, although the counterterms in eq.~\ref{expandAdS} will be more complicated.}) \cite{HS}
\beq\label{flatIR}
\PIR^{(0)}[\ta] = \exp \left\{\frac{1}{\kappa^2 R} \int d^2x \shg \, \ta^2\right\}
\eeq
As we take the limit $\ta\to \aUV\to \infty$, $\PIR$ approaches the partition function of the dual field theory, and the bare term \eqref{flatIR}\ diverges.  In order to preserve conformal invariance, we must introduce a local counterterm Lagrangian which precisely cancels this, a special case of the counterterm Lagrangian derived in \cite{Trefs}.  
From that, we see immediately from \eqref{Tcorr}\ that correlators of $T$ vanish, since after including the counterterm $\PIR$ is independent of $\ta$.          

Next, let us compute the IR amplitude for the other cases of interest and do a similar analysis. 
For $AdS_3/dS_{2}$ we find
\beq\label{AdSdSIR}
\PIR[\ta] = \exp \left\{ \frac{1}{\kappa^2 R} \int d^2x\shg \, \left( \ta \sqrt{1+\ta^2} + \text{arcsinh}\, \ta\right) -{\mc S}_{CT}[\t a] \right\} \,.
\eeq
Since the dual CFT lives on $dS_2$, it should have $T_{QFT}=\frac{c}{24\pi} {\cR}_{dS_2}$ where $c$ is the central charge, and the relation between $T$ and $T_{QFT}$ is specified in \eqref{QFTrelns}.    
Note that for large $\ta\to \aUV$,  \eqref{AdSdSIR}\ can be expanded as
\beq\label{expandAdS}
\PIR[\ta] = \exp \left\{ \frac{1}{\kappa^2 R} \int d^2x\shg \, \left( \ta^2 + \log(2\ta) + \frac{1}{2}+\dots\right) -{\mc S}_{CT}[\t a] \right\} \,,
\eeq  
where `$\dots$' represents terms of order $1/\ta^2$ and higher.  From the Poincar\'e case just covered, we know that ${\mc S}_{CT}$ contains a term cancelling the $\ta^2$ divergence.  In fact, aside from the logarithmic term, each term in the series can be cancelled by a choice of ${\mc S}_{CT}$ which takes the form of an expansion in local curvature invariants divided by powers of $\mu_L$.  The logarithmic term gives rise to the standard conformal anomaly.      

Now for $dS_3/dS_{2}$ we obtain
\beq\label{dSdSIR}
\PIR[\ta] = \exp \left\{ \frac{1}{\kappa^2 R} \int d^2x\shg \,\left( \ta \sqrt{1-\ta^2} + \arcsin \ta\right)- {\mc S}_{CT}[\t a]\right\} \,.
\eeq
In order to proceed, we need to understand ${\mc S}_{CT}$.  In the de Sitter static patch, the full geometry contains two warped throats \eqref{dSdSmet}, and dynamical $d=2$ gravity (Liouville theory).  The two matter sectors are isomorphic, but a nonzero counterterm action would break that symmetry: the path integral for one side would be multiplied by $\exp(+{\mc S}_{CT})$, while that for the other side would be multiplied by $\exp(-{\mc S}_{CT})$.  The only consistent possibility is then
\beq
{\mc S}_{CT}=0 \qquad \text{(dS case)} \,.
\eeq         
Given that, we see immediately from \eqref{dSdSIR}\ that as we go to the UV slice, $\ta\to 1$, the first derivative of $\PIR$ with respect to $\ta$ approaches zero, and we obtain
\beq\label{TdS}
\la T\ra=0 \qquad  \text{(dS case)} \,.
\eeq   
This is exactly what is expected for matter coupled to Liouville gravity:  overall this is a CFT with zero central charge.  This result was obtained in several different ways previously in \cite{dSdS, FRW}\footnote{We stress once more that we are studying properties of the matter sector coupled to gravity in the dual, and while $D-1$ dimensional gravity appears here as a constraint on this theory, we are not considering the full dynamics of lower-dimensional gravity, and the cutoff scale $\Lambda \sim 1/R$ is far below the $D-1$ dimensional Planck scale.}.  In Appendix \ref{app:3d} we will calculate $\PUV$ and the Wilsonian action for this simple case\footnote{because we can}.

\section{General framework and Hamilton-Jacobi analysis}\label{sec:framework}

The preceding sections established the main consequences of moduli stabilization for the RG evolution of the holographic dual by performing the bulk path integral. In this and the next sections we will study an equivalent framework based on the Hamilton-Jacobi formulation. This approach turns out to be helpful for deriving $\beta$ functions that include multitrace couplings and backreaction. The formulation also makes contact with QFT language, and it potentially suggests a new way of organizing the RG of the dual. 

In this section we will extend the framework for holographic Wilsonian RG of~\cite{HP}
to general curved slicings of a maximally symmetric $D$-dimensional space. The metric is chosen to be of the form
\beq\label{metricHJ}
ds_D^2=dy^2 + g_{\mu\nu}(y,x) dx^\mu dx^\nu \,,
\qquad g_{\mu\nu}(y,x)=a(y,x)^2 \hg_{\mu\nu}
\eeq
where $\hg$ is the metric of the $d$-dimensional slice ($d\equiv D-1$). This general structure will be applied to $AdS_D$ and $dS_D$ bulk geometries with different warp factors and slicings in \S \ref{sec:appl}. It will be useful to study the RG evolution of both the UV partition function $\PUV$ and the Wilsonian action introduced in (\ref{WilsonI}); a similar situation arises for functional RG in QFTs, which can be formulated in terms of the microscopic action or the 1PI effective action~\cite{functionalRG}.

\subsection{Scalar on a fixed background}\label{subsec:scalar-fixedbck}

We begin by considering a scalar field $\phi(y,x)$ in a fixed background \eqref{metricHJ} with $a(y,x)=\ba(y)$, neglecting the backreaction of the scalar field; backreaction on the metric will be taken into account below. 

In order to determine the radial evolution of the UV amplitude, it is useful to begin from the bare version of the amplitude $\PUV^{(0)}$; recall from \S\ref{sec:results} that the two are related by
\be\label{eq:PUVCT}
\PUV(\tphi,\ta,\l) =  e^{\kappa^{-2} \cS_{CT}[\t\phi,\t a]} \PUV^{(0)} e^{-\kappa^{-2} \cS_{CT}[\phiUV,\aUV]} \,
\ee
where $\cS_{CT}=\int d^dx \sqrt{g} {\cal L}_{CT}$. Here $\t a = \bar a(L)$ because the background is fixed; later we will include backreaction of the scalar on $\t a$. For a scalar field with Euclidean action
\beq\label{action}
\cS^{(0)}= \int dy d^dx \, \ba^d \shg \left\{ \frac{1}{2} \left(\pd{\phi}{y}\right)^2 + \frac{1}{2\ba^2} \hg^{\mu\nu} \p_\mu\phi \p_\nu\phi + V(\phi) - \frac{1}{2} \bar{\mc R}^{(D)}\right\} \,,
\eeq
the radial evolution is governed by a radial Schr\"odinger equation
\be\label{PUVSchr0}
\kappa^2 \p_\l \PUV^{(0)}(\tphi,L)= H^{(0)}(\tphi,\tpi_\phi) \PUV^{(0)}(\tphi,L)
\ee
with the Hamiltonian\footnote{Our definition of momentum in Euclidean signature is $\tpi_\phi = i \frac{1}{\sqrt{\t g}} \frac{\partial{\cal L}}{\partial \tphi_{,y}}$ and $H =\int d^d x \sqrt{\t g}( i\sqrt{\t g}\,\tpi_\phi \tphi_{,y} + {\cal L})$.  Quantizing the theory according to $[\tphi(x),\tpi_\phi(x')] = i \kappa^2 \frac{\delta^{(d)}(x-x')}{\sqrt{\t g}}$ then gives $\tpi_\phi = -i \kappa^2 \frac{1}{\sqrt{\t g}}\frac{\delta}{\delta \tphi}$.} 
\beq\label{Ham}
H^{(0)}(\tphi,\tpi_\phi) = \int d^dx \sqrt{\t g}\left( \frac{1}{2}\tpi_\phi^2+ \frac{1}{2} \t g^{\mu\nu} \p_\mu \tphi \p_\nu \tphi + V(\tphi) -\frac{1}{2} \bar{\mc R}^{(D)}  \right)\,.
\eeq
Here $\t \pi_\phi= - i \kappa^2 \frac{1}{\sqrt{\t g}} \frac{\delta}{\delta \t \phi}$ is the (covariant) canonical momentum, $\t g_{\mu\nu} = \t a^2 \hat g_{\mu\nu}$, and $\bar{\mc R}^{(D)}$ is the background $D$-dimensional Ricci scalar.

Given this, the radial evolution of the UV amplitude including counterterms becomes
\be\label{PUVSchr}
\kappa^2 \p_\l \PUV(\tphi,L)= H(\tphi,\tpi_\phi) \PUV(\tphi,L)
\ee
with the new Hamiltonian
\be\label{eq:Hfinite1}
H = e^{\kappa^{-2} \t \cS_{CT}} H^{(0)} e^{-\kappa^{-2} \t \cS_{CT}}+ \partial_L \t \cS_{CT}\,.
\ee
Note that only the counterterm $\t \cS_{CT} \equiv \cS_{CT}(\t\phi,\t a)$ contributes to the evolution of the amplitude; $\cS_{CT}(\phiUV,\aUV)$ cancels out from this equation. Also, $\partial_L \t \cS_{CT}$ appears because the background $\t a=\ba(\l)$ contains explicit $L$ dependence. This term will be absent once we include backreaction on the scale factor. 

Commuting the $e^{-\kappa^{-2} \t \cS_{CT}}$ factor to the left of $H^{(0)}$ has the effect of replacing $\tpi_\phi$ by $\tpi_\phi + \frac{i}{\sqrt{\t g}} \frac{\delta \t \cS_{CT}}{\delta \tphi}$ in \eqref{Ham}.  Therefore the new Hamiltonian \eqref{eq:Hfinite1} becomes
\begin{multline}\label{eq:Hfinite2}
H(\t \phi, \t \pi_\phi)=\int d^d x \sqrt{\t g}\, \Big[\frac{1}{2}\Big(\t \pi_\phi +  \frac{i}{\sqrt{\t g}} \frac{\delta \t \cS_{CT}}{\delta \t \phi}\Big)^2+ \frac{1}{2} \t g^{\mu\nu} \p_\mu \t \phi \p_\nu \t \phi + V(\t \phi) - \frac{1}{2} {\bar {\mc R}}^{(D)}  \\
+(\p_\l + d \partial_L \log \t a) \t {\mc L}_{CT}\Big]\,.
\end{multline}
Writing the UV amplitude as
\beq\label{WUV2}
\PUV(\tphi,\l)= \exp\left(-\kappa^{-2} \WUV(\tphi,\l)\right) \,,\quad
\WUV(\tphi,\l)= \int d^dx \, \sqrt{\t g}\, w(\tphi,\l) \,,
\eeq
and taking the semiclassical limit $\kappa^2 \to 0$, the Schr\"odinger equation \eqref{PUVSchr} becomes a Hamilton-Jacobi (HJ) equation for $\WUV$
\beq\label{eq:HJWUV}
\p_\l \WUV = -H\left(\tphi,\frac{i}{\sqrt{\t g}}\frac{\delta\WUV}{\delta\tphi}\right) \,.
\eeq
If we restrict to the zero mode of $\tphi$ on the $d$-dimensional slice, the HJ equation for $w$ can be further simplified to
\beq\label{wHJ}
(\p_\l + d \partial_L \log \t a)(w+V_{CT}) = \frac{1}{2} \left(\frac{\partial}{\partial\tphi}(w+V_{CT})\right)^2 - V(\tphi)+\frac{1}{2} {\t {\mc R}}^{(D)}\,.
\eeq
Here $V_{CT}$ is the zero-derivative term in $\mc L_{CT}$. Thus, $w+V_{CT}$ satisfies the same HJ equation as the bare $w^{(0)}$, consistent with (\ref{eq:PUVCT}) and the fact that the counterterm at $\LUV$ does not contribute to the radial evolution.

The Wilsonian action $s(\cO,\l)$ is determined in terms of the integral transform (\ref{WilsonI}).  For a scalar field on a fixed background, the integral transform is simply
\beq\label{intTrans}
\exp\left(-\kappa^{-2} s(\cO,\l)\right)= \int \cD \tphi\, \PUV(\tphi, \l) \, e^{\kappa^{-2} \int d^dx \sqrt{\t g}\, \tphi \cO} \,.
\eeq
Using this and (\ref{PUVSchr}), we can also derive an equation governing the radial evolution of $s$ directly:
\beq\label{sSchr}
\kappa^2 \p_\l e^{-\kappa^{-2} s(\cO,\l)} = \left[H\left(\frac{\kappa^2}{\sqrt{\t g}} \frac{\delta}{\delta\cO}, i \cO \right) + (d \partial_L\log \t a) \int d^dx \, \cO \kappa^2\frac{\delta}{\delta\cO}\right] e^{-\kappa^{-2} s(\cO,\l)} \,,
\eeq
with the extra term on the right hand side coming from the derivative acting on $\sqrt{\t g}$ in \eqref{intTrans}. Restricting as before to the zero mode and taking the semiclassical limit $\kappa^2 \to 0$, the Wilsonian Lagrangian $\sigma$ defined by
\beq\label{sigdef}
s(\cO,\l)= \int d^dx \, \sqrt{\t g}\, \sigma(\mc O,\l)
\eeq
evolves according to
\begin{multline}\label{sigHJ}
\p_\l \sigma + d \partial_L \log \t a \left(\sigma-\cO\pd{\sigma}{\cO}\right) = \frac{1}{2} \left[\cO+ V'_{CT}\left(-\frac{\partial \sigma}{\partial \mc O}\right) \right]^2 - V\left(-\pd{\sigma}{\cO}\right)+ \frac{1}{2}  {\t {\mc R}}^{(D)}\\
- (\partial_L + d \partial_L \log \t a) V_{CT}\left(-\frac{\partial \sigma}{\partial \mc O}\right) \,.
\end{multline}

Of course, in the semiclassical limit the path integral $\PUV$ can be evaluated directly by plugging in the classical solution between $\LUV$ and $L$ with the appropriate boundary conditions, and $\sigma$ follows from the integral transform (\ref{WilsonI}). This was the more direct approach already used in \S\S \ref{sec:results} and \ref{sec:3d}.

\subsection{Including backreaction}\label{subsec:dyn}

Having obtained the RG equation for the Wilsonian action of a single operator $\mc O$ (and its multi-trace deformation), let us now add backreaction on the metric. The metric $g_{\mu\nu}$ is dual to the QFT stress tensor $T^{\mu\nu}$, so allowing for a dynamical metric amounts to taking into account the effects of interactions between $\mc O$ and $T^{\mu\nu}$. For simplicity we restrict to a minisuperspace analysis, and only analyze the scale factor $a$ dual to the trace of the stress tensor $T$.

As before, the radial evolution for $\PUV$ is obtained from that of $\PUV^{(0)}$ after taking into account the effects of the counterterm $\t {\mc S}_{CT}$. Setting the lapse to one and the shift to zero, the bare ADM Hamiltonian for the scale factor and scalar field is
\be\label{H0back}
H^{(0)}(\t \phi_I, \t \pi_I) = \int d^dx \sqrt{\t g}\left( -\frac{1}{2d(d-1)}\t a^2 \t \pi_a^2 + \frac{1}{2} \t \pi_\phi^2+ \frac{1}{2} \t g^{\mu\nu} \partial_\mu \t \phi \partial_\nu \t \phi + V(\t \phi) - \frac{1}{2} \t{\mc R}^{(d)}\right)\,,
\ee
where $\t g_{\mu\nu}(x) = \t a(x) \hat g_{\mu\nu}(x)$ is the induced $d$-dimensional metric, and $\t{\mc R}^{(d)}$ is its Ricci scalar containing the last two terms in \eqref{Rprime}. We also use $\t \phi_I$ to denote $\t \phi$ and  $\t a$. The canonical momenta are given by $\t \pi_I=-i \kappa^2 \frac{1}{\sqrt{\t g}} \frac{\delta }{\delta \t \phi^I}$. As discussed in~\cite{HP}, the Hamiltonian constraint should not be imposed on $\PUV^{(0)}$.

Writing $\PUV^{(0)}$ in terms of $\PUV$ and the counterterms leads to the evolution equation
\be\label{PUVSchr-back}
\kappa^2 \p_\l \PUV(\tphi_I,L)= H(\tphi_I,\tpi_I) \PUV(\tphi,L)\,,\quad
H = e^{\kappa^{-2} \t \cS_{CT}} H^{(0)} e^{-\kappa^{-2} \t \cS_{CT}}\,.
\ee
Note that $ \t \cS_{CT}= \cS_{CT}(\t \phi_I, \t a)$ is independent of $L$ because $\t a$ is taken to be dynamical -- the term $\partial_L \t \cS_{CT}$ found before in (\ref{eq:Hfinite1}) is now absent. From this point, the derivation continues analogously as in \S \ref{subsec:scalar-fixedbck}.  The effect of the counterterm is to shift the canonical momenta,
\bea\label{eq:Hfinite3}
H(\t \phi_I, \t \pi_I)&=&\int d^d x \sqrt{\t g}\, \Big[-\frac{1}{2d(d-1)}\t a^2 \left(\t \pi_a+ \frac{i}{\sqrt{\t g}} \frac{\delta \t \cS_{CT}}{\delta \t a}\right)^2 +\frac{1}{2}\left(\t \pi_\phi +  \frac{i}{\sqrt{\t g}} \frac{\delta \t \cS_{CT}}{\delta \t \phi}\right)^2  \nonumber\\
&+&\frac{1}{2} \t g^{\mu\nu} \p_\mu \t \phi \p_\nu \t \phi + V(\t \phi) - \frac{1}{2} {\t {\mc R}}^{(d)}\Big]\,.
\eea
The radial evolution of the UV free energy $\WUV=-\kappa^2 \log\PUV$ is then given in the semiclassical limit $\kappa^2 \to 0$ by the HJ equation
\beq
\p_\l \WUV= -H\left(\tphi_I, i \frac{1}{\sqrt{\t g}}\frac{\delta\WUV}{\delta\tphi_I} \right) \,,
\eeq
where the ordering ambiguity between $\ta$ and $\tpi_a$ in \eqref{H0back} and \eqref{eq:Hfinite3} goes away.
This implies that $\WUV+\t \cS_{CT}$ evolves with the bare hamiltonian $H^{(0)}$, as expected.

The dependence on the zero modes of $\t a$ and $\t \phi$ can be more easily analyzed in terms of the density $w$ defined as
\beq
\WUV( \ta,\tphi,\l)= \int d^dx \, \sqrt{\t g} \, w(\ta,\tphi,\l) \,.
\eeq
The evolution equation for $w+V_{CT}$ is then the same as that of the bare $w^{(0)}$,
\beq\label{wHJa}
\p_\l (w+V_{CT}) =
-\frac{1}{2d(d-1)} \t a^2\left(\frac{\partial}{\partial \t a}(w+V_{CT})\right)^2
+\frac{1}{2} \left(\frac{\partial}{\partial \t \phi}(w+V_{CT})\right)^2
-V(\tphi) + \frac{1}{2\t a^2} \hat {\mc R} \,,
\eeq
where we have used $\t {\mc R}^{(d)}=\hR/\ta^2$ at the level of zero mode analysis.
\eqref{wHJa} is the generalization of \eqref{wHJ} for the case of a dynamical scale factor. (For this reason, in this expression there is no term proportional to $\partial_L \log \t a$; we also added $\partial_L V_{CT}=0$ to the left hand side, to have the equation in terms of $w+V_{CT}$).

In the presence of a dynamical scale factor, the Wilsonian action depends on both $\mc O$ and the trace of the stress tensor $T$. At the linear order, the coupling between the metric and stress tensor on a surface $y=L$ is
\be
\frac{1}{2}\int d^dx\,\sqrt{\t g}\, \delta \t g_{\mu\nu}\,T^{\mu\nu} + \ldots =  \int  d^dx\,\sqrt{\t g} \left(\frac{\delta \t a}{\t a}+ \ldots\right) \,(\t g_{\mu\nu}T^{\mu\nu})\,.
\ee
The `$\ldots$' are higher order terms that are needed such that the usual relation 
$$
T_{\mu\nu}= \frac{2}{\sqrt{g_{UV}}} \frac{\delta \log Z}{\delta g^{\mu\nu}_{UV}}
$$ 
is obtained.
In all, the Wilsonian Lagrangian $\sigma(T, \mc O,L)$ is given in terms of $w(\t a, \t \phi, L)$ by
\be\label{eq:stransf}
\exp\left(- \frac{1}{\kappa^2} \int d^dx\,\sqrt{\bar{g}(L)}\, \sigma \right) = \int \mc D \t\phi \mc D \t a\,\exp\left(-\frac{1}{\kappa^2} \int d^dx\,\sqrt{\t g} \left[ w - \left(\frac{\delta \t a}{\t a}+ \ldots \right) T - \t \phi \mc O\right] \right)\,.
\ee

This is the functional that should be compared with the Wilsonian action of the QFT dual, for the case with single- and multi-trace interactions for both $\mc O$ and the trace of the stress tensor $T= \t g_{\mu\nu} T^{\mu\nu}$. As before, instead of performing the integral transform of $w$, one can derive a HJ equation for $\sigma$ itself. 

\section{Moduli stabilization and holography for (A)dS$_D$}\label{sec:appl}

We are now ready to apply the general techniques developed in \S\ref{sec:framework} to maximally symmetric $AdS_D$ and $dS_D$ solutions. Our goal is to determine the consequences of moduli stabilization $V'(\phi_*)=0$ for the structure of the interactions in the holographic dual. We will focus on the functional $\WUV(\t a, \t \phi, L)$, whose HJ equation is simpler than that of the Wilsonian action.

\subsection{RG flow in the zero mode sector}\label{subsec:RG0m}

Before studying the full problem, as a warmup let us first analyze the RG flow restricted to the zero mode sector.  For simplicity in this section we will shift $\phi$ such that the minimum of the potential occurs at $\phi=0$; i.e.\ we take $\phi_*=0$. 

Ignoring backreaction on the metric, the RG equation is given by (\ref{wHJ}). While it is hard to find exact solutions, we will work around the minimum of the potential and find a series solution in powers of $\t \phi$. Furthermore, the counterterm potential $V_{CT}$ will be chosen to respect the critical point, $V'_{CT}(0)=0$. We first solve the HJ equation for $\hat w(\t \phi, L) \equiv w(\t \phi, L)+ V_{CT}(\t \phi, L)$ and then translate to $w$.

In terms of the series
\beq\label{Vwexp}
V(\tphi)=\sum_{n=0}^{\infty} \frac{1}{n!} V^{(n)}_* \tphi^n \,,\quad
\hat w(\tphi,\l)=\sum_{n=0}^{\infty} \frac{1}{n!} \hat w_n(\l) \tphi^n \,,
\eeq
we may organize the HJ equation \eqref{wHJ} order by order as
\begin{align}
(\p_\l+ d \partial_L \log \t a) \hat w_0 &= \frac{1}{2} \hat w_1^2 - V_* \,,\\
\label{w1}
(\p_\l+ d \partial_L \log \t a) \hat w_1 &= \hat w_1 \hat w_2 - V'_* \,,\\
(\p_\l+ d \partial_L \log \t a) \hat w_2 &= \hat w_2^2 +\hat w_1 \hat w_3 - V''_* \,,\\
&\cdots\nn\\
\label{wn}
(\p_\l+ d \partial_L \log \t a)\hat w_n &= \frac{1}{2} \sum_{m=0}^n \frac{n!}{m!(n-m)!} \hat w_{m+1} \hat w_{n-m+1} - V^{(n)}_* \,,
\end{align}
where $V^{(n)}_* \equiv V^{(n)}(0)$.

These general results are valid as long as $\t \phi$ remains small. We now impose that the modulus is stabilized at $\t \phi =0$, namely $V'(0)=0$.
This consistently allows $\hat w_1=0$ to be a solution to \eqref{w1}, with any maximally symmetric $d$-dimensional slicing.  The HJ equations can then be solved iteratively, with the equation involving $\p_\l \hat w_n$ depending only on coefficients $\hat w_m$ with $m\leq n$. Choosing the counterterm such that $V'_{CT}(0)=0$, the same properties apply to the generating function $w(\t \phi, L)$.

For $AdS_D$ in flat slicing, we have a stronger statement that the function $w(\tphi,\l)$ is actually independent of $\l$.  This is because the scale factor $\t a=e^L$ gives $\partial_L \log \t a=1$, allowing us to set the right hand side of the HJ equations \eqref{wn} to zero and solve for $\hat w_n$ (and hence for $w_n$) algebraically.  The first few coefficients in this case are
\beq\label{wflat}
\hat w_0 = -\frac{V_*}{d} \,,\quad
\hat w_2 = \Delta_\pm \equiv \frac{d}{2} \pm \sqrt{\frac{d^2}{4}+ V''_*} \,,\quad
\hat w_3 = \frac{V^{(3)}_*}{3\Delta_{\pm}-d} \,,\quad\cdots
\eeq
However, for $(A)dS_D$ with $dS_{D-1}$ slices it is no longer consistent to impose $\p_\l w=0$ because $\partial_L \log\t a$ is $\l$-dependent. In the $AdS_D/dS_{D-1}$ case, this is explained by noting that the dual CFT lives on a curved background $dS_{D-1}$, which breaks explicitly the conformal invariance.\footnote{Note that as is familiar from AdS/CFT, reality of operator dimensions and of $\hat w_2$ implies that the scalar potential satisfies the Breitenlohner-Freedman bound\cite{BF}.  In the de Sitter case, we require $V'' > 0$ to avoid perturbative instability on the gravity side, and since the $d$-dimensional couplings are dynamical any instability in the bulk will translate to an instability on the $d$-dimensional side.} Nevertheless, we can still set $w_1=0$ and solve \eqref{wn} iteratively.

The Wilsonian action can be derived from the integral transform of $w$ or from \eqref{sigHJ}.
The expansion in powers of $\t \phi$ corresponds to an expansion in powers of the dual operator $\mc O$
\beq
\sigma(\cO,\l)=\sum_{n=0}^{\infty} \frac{1}{n!} \sigma_n(\l) \cO^n \,,
\eeq
where $\sigma_n$ is identified with the $n$-th multi-trace coupling at a certain cutoff scale $\mu_L$ in the holographic dual. The HJ equation shows that if $V'(0)=0$, there is a constant solution $\sigma_1=0$ and, precisely in this case, the dependence of $\partial_L \sigma_n$ on higher $\sigma_{m>n}$ cancels. This allows for an iterative solution, where the equation for $\p_\l \sigma_n$ depends only on $\sigma_{m \leq n}$. 

Equivalently, as explained in \S\ref{sec:results}, in the semiclassical limit the integral transform \eqref{intTrans} is dominated by the saddle point solution $\cO=\p w/\p\tphi$. The point is that since $w_1=0$, there is an order by order solution that starts linear in $\mc O$,
\beq
\t\phi =\frac{1}{w_2}\cO -\frac{w_3}{2w_2^3}\cO^2 +\cdots \,,
\eeq
and $\t \phi\propto \mc O$ at leading order implies that the single trace coupling $\sigma_1 = 0$. The first few orders relating $\sigma$ to the $w_n$ are then
\beq\label{sigw}
\sigma(\cO,\l) =w_0(\l) -\frac{1}{2w_2(\l)}\cO^2 +\frac{w_3(\l)}{6w_2^3(\l)}\cO^3 +\cdots
\eeq
As before, except for the case of Poincar\'e AdS, only the single-trace coupling does not run, with the multi-trace ones having a nontrivial dependence on the RG scale $\mu_L$. 

This analysis was in the limit of a fixed background. Backreaction on the metric can be incorporated by making $w_n$ a function of $\t a$:
\beq
w(\ta,\tphi,\l)=\sum_{n=0}^{\infty} \frac{1}{n!} w_n(\ta,\l) \tphi^n \,.
\eeq
and then solving (\ref{wHJa}). We will give more details of the case with dynamical metric below, and here just note that if the modulus is stabilized, there is still a consistent solution with $w_1=0$. Choosing this then implies that the equation  for $\partial_L w_n$ (which is now a PDE) only depends on the lower $w_{m \leq n}$. Similar results apply to the Wilsonian action, where now the coefficients $\sigma_n$ are functions of $T$ and $L$.

\subsection{Wilsonian generating functional}\label{subsec:WuvRG}

Having understood the RG flow restricted to the zero mode sector, let us now discuss the full functional $\WUV$ including sources for $\mc O$ and for the trace of the stress tensor.\footnote{A source for the traceless part of the stress tensor can be treated similarly to the source for $\mc O$, by adding a term $\left(\frac{1}{\sqrt{\t g}} \frac{\delta {\hWUV}}{\delta \t h_{\mu\nu}(x)} \right)^2$ to the HJ equation, where $\t h_{\mu\nu}$ is a traceless metric fluctuation.} It satisfies a Callan-Symanzik type equation
\be\label{eq:CS}
\partial_L \hWUV = \int_x \left[- \frac{1}{2d(d-1)}  \left(\frac{\t a}{\sqrt{\t g}} \frac{\delta \hWUV}{\delta \t a} \right)^2 + \frac{1}{2}\left(\frac{1}{\sqrt{\t g}} \frac{\delta \hWUV}{\delta \t \phi} \right)^2- \frac{1}{2} \t g^{\mu\nu} \partial_\mu \t \phi \partial_\nu \t \phi - V(\t \phi) + \frac{\t {\mc R}^{(d)}}{2}  \right]
\ee
where we have introduced the combination $\hWUV \equiv \WUV + \t {\mc S}_{CT}$, and the integral symbol contains the volume factor $d^dx \sqrt{\t g}$.

We will solve this equation in an expansion in powers of $\t \phi(x)$ around the minimum of the potential:
\be
\WUV(\t \phi, \t a, L)= W^{(0)}(\t a, L) + \int_x W^{(1)}(x;\t a, L) \t \phi(x) + \frac{1}{2} \int_{x_1, x_2}W^{(2)}(x_1,x_2;\t a, L) \t \phi(x_1) \t \phi(x_2) + \ldots\,,
\ee
with a similar expansion for $\t {\mc S}_{CT}$. Here, each coefficient $W^{(n)}$ is a functional of the dynamical scale factor. The equation satisfied by $W^{(n)}$ is obtained by taking the $n$th functional derivative of (\ref{eq:CS}) with respect to $\t \phi$, and then evaluating the expression at $\t \phi=0$. To avoid cluttering our formulas, in what follows the dependence on $\t a$ and $L$ is kept implicit.

At zeroth order, we find
\be
\partial_L \hat W^{(0)} = \int_x \left[- \frac{1}{2d(d-1)} (D_{\t a(x)} \hat W^{(0)})^2 + \frac{1}{2} \hat W^{(1)}(x)^2- V(0) + \frac{1}{2} \t {\mc R}^{(d)} \right]\,,
\ee
with the shorthand notation $D_{\t a(x)} \equiv \frac{\t a}{\sqrt{\t g}} \frac{\delta }{\delta \t a}$. The first order coefficient, which determines the running of the single trace operator, satisfies the equation
\be\label{eq:W1}
\partial_L \hat W^{(1)}(x_1) = \int_x \left[- \frac{1}{d(d-1)}  D_{\t a(x)} \hat W^{(0)}\, D_{\t a(x)} \hat W^{(1)}(x_1) + \hat W^{(1)}(x) \hat W^{(2)}(x,x_1) \right] - V'(0)\,.
\ee
At second order,
\bea
\partial_L \hat W^{(2)}(x_1,x_2) &=& \int_x \Big[- \frac{1}{d(d-1)}  (D_{\t a(x)} \hat W^{(1)}(x_1)\, D_{\t a(x)} \hat W^{(1)}(x_2)+ D_{\t a(x)} \hat W^{(1)}(x) D_{\t a(x)} \hat W^{(2)}(x_1,x_2)) \nonumber\\
&+& \hat W^{(2)}(x,x_1) \hat W^{(2)}(x,x_2) +
 \hat W^{(1)}(x) \hat W^{(3)} (x,x_1,x_2) \Big] + (\t \Box_{x_1}- V''(0)) \delta^d(x_1-x_2)\nonumber
\eea
and similarly for the higher order coefficients.

Now we impose the condition that the modulus is stabilized. Then, (\ref{eq:W1}) is solved by $\hat W^{(1)}(x; \t a, L)=0$. Choosing the counterterm to preserve this critical point ($\t {\mc S}_{CT}^{(1)}=0$), we deduce that the first order coefficient $W^{(1)}=\hat W^{(1)}-\t {\mc S}_{CT}^{(1)}$ of the generating functional vanishes identically. In this case, the Callan-Symanzik  equation for $W^{(2)}$ depends only on $W^{(0)}$ and $W^{(1)}$. This iterative structure continues to hold for the higher $W^{(n)}$.

For stabilized moduli, computing the Wilsonian action as the integral transform of $\WUV$ shows that the single-trace couplings in the dual QFT do not run, and that higher-trace couplings only depend on lower-trace ones, with a specific $L$ dependence that encodes the simple shape of the bulk warp factor. This implies important simplifications for the Wilsonian RG flow of the QFT, suggesting a new way of organizing the path integral.

\section{Metastability}\label{sec:meta}

In this section we briefly consider the implications of the metastability of $V(\phi)$ in cases such as dS (and a subset of perturbative AdS solutions) where this occurs.
The basic effect it has for the holographic RG is to introduce imaginary contributions into the Wilsonian action.  This seems reasonable, reflecting the fact that the theory is not unitary by itself if it decays -- more degrees of freedom are required to capture the decay product \cite{FRW, HarlowSusskind}.\footnote{Another effect of the metastability of $V(\phi)$ is simply to introduce CdL decays in the $D-1$ gravity + matter dual theory to $dS_D$, but here we are focusing on what it says about the holographic Wilsonian effective action in each throat QFT.}  It provides an interesting new application of basic instanton technology to the holographic RG.    

For simplicity let us analyze the scalar on a fixed $dS_D$ background; we expect the backreacted case to work similarly as happened in the perturbative analysis in the previous sections.
Formally $\PUV$, which determines the Wilsonian action, is like a transition amplitude
\be\label{Psitrans}
\PUV[\tilde\phi,L]=\langle \phi_* | {\mc R}(e^{-\int_{\l}^{\LUV} H}) |\tphi \rangle
\ee
satisfying the radial Schr\"odinger equation.  Here we put the UV value of $\phi$ at the local minimum, $\phiUV=\phi_*$, and $\mc R$ denotes radial-coordinate ordering.

We will continue to use a semiclassical ($\kappa^2\to 0$) limit to control the calculations.  There are contributions to $\PUV$ from classical solutions to the $\phi$ field equation which start at $\phi_*$ at the UV slice and end at $\tilde\phi$ at the slice $y=L$, times a determinant $K$ from integrating over the fluctuations about each solution.  Since we are doing radial rather than time evolution, the equation for the mode of $\phi$ which is homogeneous along the $dS_{D-1}$ is equivalent to the equation for a particle rolling on the inverse of our potential, $-V(\phi)$, with a friction term from the warp factor.  

This is isomorphic to the equation governing instanton solutions for fields on $dS_D$, but here the interpretation will be different.    In instanton physics, one formally computes the imaginary part of the ground state energy and infers from it a decay rate.  In our case, a somewhat similar computation should give the imaginary part of $\WUV$ and the Wilsonian action:\  schematically
\be\label{structure}
\PUV[\tilde\phi,L]=e^{-W_{\rm UV, pert}[\tilde\phi,L]}+\sum e^{-W_{\rm inst}[\tilde\phi,L]}K[\tilde\phi,L]
\ee  
with the determinant $K$ being imaginary for fluctuations about a single instanton solution.  
As for instantons, we will be justified in considering non-perturbatively small contributions without first summing all perturbative effects because the nonperturbative effects will be the leading contribution to the imaginary part of the action.         

We are in the false vacuum in the UV. At the slice $y=L$ we could be in the basin of attraction of the false or true vacuum, or right at the extremum between them, depending on $\tilde\phi$.  At the perturbative level we focused on the behavior near the local minimum $\phi_*$, where $\PUV$ is approximately Gaussian.  To start we could consider that regime here, meaning that at slice $y=L$ the field is in the basin of attraction of the original false vacuum.   In addition to the original solution $\phi=\phi_*$, there can be bounce solutions in which the field goes to the other basin and comes back, in general doing so multiple times.  In particular, we could consider for simplicity the case that the potential admits bounce solutions which are thin compared to the curvature scale of the $dS_D$, which reduces the problem to simple particle mechanics on the inverted potential without a significant friction term, the simplest case in \cite{Coleman}.  For $\tilde\phi=\phi_*$, the action for such solutions will be the standard bounce action, with imaginary $K$ from the negative mode.  
For our purposes, we are also interested in more general configurations $\tilde\phi\ne\phi_*$, and we expect these also generate a complex effective theory at a nonperturbatively suppressed level.       

\section{Summary and future directions}\label{sec:concl}

In the present work we have related the statement that all moduli are stabilized in $(A)dS_D$ to the property that single-trace couplings in the Wilsonian effective action for the holographic dual theory living on $dS_{D-1}$ have vanishing $\beta$ functions.  This result applies equally well to perturbatively stable de Sitter and anti-de Sitter solutions.  In both cases the RG also exhibits the simplifying feature that multiple trace terms are determined by lower trace ones.   It extends also to cases where moduli stabilization is incomplete:  there the single-trace $\beta$ functions vanish for each operator dual to a perturbatively stabilized direction. In the metastable case with non-perturbative decays, the Wilsonian action exhibits the expected breakdown in unitarity at a non-perturbatively suppressed level, reflecting the need for additional degrees of freedom to capture its decay to a more general FRW background.  

Further, in analyzing the correlators of the trace $T$ of the stress-energy tensor, we found 
that the joining of the two warped throats comprising the causal patch of de Sitter fixes the (otherwise arbitrary) counterterm Lagrangian to zero, leading to a clean result that $T=0$ in the de Sitter case.  This is exactly what is expected from the coupling of the matter sector to dynamical gravity (Liouville in $d=D-1=2$), agreeing with earlier calculations \cite{FRW}\cite{dSdS}.     

These features and others developed in \cite{dSdS}\cite{dSdSbrane}\ hang together, indicating a consistent framework for de Sitter spacetime.  The vanishing of the single-trace beta functions should provide strong guidance in developing concrete dual theories further.   It will be interesting to understand this from a more microscopic perspective in simple brane constructions which uplift the potential and metastabilize one or more of the moduli \cite{inprogress}.  We would also like to understand it in the concrete brane construction in \cite{dSdSbrane}\ and to use it as a strong constraint on the content and couplings of new examples of de Sitter duals.  Additional macroscopic calculations, such as correlation functions in each throat theory, also help constrain the duals and may define them in the large-$N$ limit, at least for the low-dimensional cases where gravity just provides a constraint.  

It also remains of interest to understand the role of the global eternally inflating geometry; see  \cite{Anninos:2011zn, Anninos:2012qw} for some recent works on this subject.   The observer patch is all that is operationally accessible, and it is possible that the global correlation functions do not contain additional information \cite{Bousso:2009mw}.  In any case, there is much to do to flesh out the holographic  (re-)construction observables in de Sitter and FRW geometries.

\section*{Acknowledgements}

We thank D. Anninos, D. Harlow, J. Polchinski, and S. Shenker for helpful discussions, as well as the organizers and participants of the KITP program ``Bits, Branes, Black Holes'' and the Bay Area de Sitter working group. GT would like to dedicate this work to Ana Sof\'ia and her mother Anabela. This work was supported in part by the National Science Foundation under grant PHY-0756174 and by the Department of Energy under contracts DE-AC03-76SF00515 and DE-FG02-92-ER40699.  

\appendix

\section{Treating the term linear in $\delta\ta$}\label{app:T0}
In this appendix, we show that the linear term $w_{01}\delta\ta$ in $\PUV$ does not produce single-trace couplings for $\cO$.  In particular, we will see that $w_{01}$ removes the unit operator part of $T$ in just the right way, so that at the saddle point solution $\delta\ta$ is replaced by (a combination of) multi-trace operators.

Let us start by writing $\WUV=\kappa^2\log\PUV$ as a double expansion
\beq
\WUV(\ta,\tphi,\l)=\int d^dx \, \ba(\l)^d \shg \sum_{m,n=0}^\infty \frac{1}{m!n!} w_{mn}(\l) (\delta\tphi)^m (\delta\ta)^n \,,
\eeq
where as before $\delta\tphi=\tphi-\phi_*$, $\delta\ta=\ta-\ba(\l)$.  Note that we have chosen to have $\ba(\l)^d$ in front of the sum, putting all $\ta$-dependence into the expansion.  To obtain the Wilsonian action, we perform an integral transform \eqref{WilsonI}:
\beq\label{WilsonI2}
\exp\left(-\kappa^{-2} s(\cO,\l)\right)= \int \cD \ta \cD \tphi\, \PUV(\tphi,\t a, \l) \, e^{\kappa^{-2} \int d^dx \sqrt{\hg} \left[\ta^d \tphi \cO +\ba(L)^{d-1} \delta\ta \, T+\ldots\right]} \,.
\eeq
The saddle point that dominates this integral can be obtained by solving
\begin{align}
\cO=&w_{20}\delta\tphi+(\text{higher orders in $\delta\tphi$, $\delta\ta$}) \,, \\
\label{saddleT}
\frac{T}{\ba(\l)}=&w_{01}+w_{02}\delta\ta+(\text{higher orders in $\delta\tphi$, $\delta\ta$}) \,,
\end{align}
where we have used the fact that there is no term linear in $\delta\tphi$ (i.e.\ $w_{10}=0$).

The trace of the stress-energy tensor is a redundant operator; it may be written in the form
\beq
T=T_0 + \beta_i \cO_i + (\text{multi-trace operators}) \,,
\eeq
where $T_0$ denotes the part of $T$ that is proportional to the unit operator, and it may be found by varying $\WIR$ with respect to $\ta$ a little away from the background solution:
\beq
T_0=-\frac{1}{\ba(\l)^{d-1}\shg} \left. \frac{\delta\WIR(\ta,\tphi,\l)}{\delta\ta} \right|_{\ta=\ba(\l),\, \tphi=\phi_*} \,.
\eeq
This is consistent with the postulate \eqref{PIRQFT} about $\PIR$, where $T_0$ can be thought of as the expectation value of $T$ when the source fluctuations $\delta\tphi$, $\delta\ta$ are turned off.

We wish to relate $T_0$ to the linear term $w_{01}$.  To achieve this, we note that at the semiclassical level, we must have
\beq
\frac{1}{Z} \int \cD \ta \cD \tphi \, \PUV \PIR \, \ta =\ba(\l) \,.
\eeq
This means that the exponent $\WUV+\WIR$ must be stationary at the background solution:
\beq
\left. \frac{\delta(\WUV+\WIR)}{\delta\ta} \right|_{\ta=\ba(\l),\, \tphi=\phi_*} =0 \,,
\eeq
which immediately leads to a relation between $T_0$ and $w_{01}$:
\beq
T_0 = w_{01} \ba(\l) \,.
\eeq
Plugging this into \eqref{saddleT}, we find that the saddle point for \eqref{WilsonI2} can be consistently solved to be
\beq
\delta\tphi=\frac{\cO}{w_{20}} +\dots\,, \qquad
\delta\ta=\frac{T-T_0}{w_{02}\ba(\l)} +\dots\,.
\eeq
Substituting this saddle point solution into \eqref{WilsonI2}, we find that the terms linear in $T-T_0$ cancel.  Since terms quadratic in $T-T_0$ or higher are multi-trace operators, they do not affect our result that the single-trace couplings have vanishing beta functions.

\section{$\PUV$ and Wilsonian Action for $(A)dS_3/X_2$}\label{app:3d}

In this appendix we continue the calculations begun in \S\ref{sec:3d}, working out $\PUV$ and the Wilsonian action for the simplest example.  

\subsection{UV amplitude}

Let us calculate the UV amplitude $\PUV$ for $(A)dS_3/X_2$ directly from \eqref{PsiUV}.  In $d=2$, the bulk action for the $d$-dimensional zeromodes without counterterms again becomes \eqref{Sa2d} at fixed $\phi=\phi_*$.
We would like to compute the path integral \eqref{PsiUV} with the boundary conditions $a(\l)=\ta$, $a(\LUV)=a_{0}$.  We will again focus on the zero mode first.  The integral is then given by the action evaluated at the unique classical trajectory (satisfying the equation of motion) which travels from $\ta$ to $a_0$ within Euclidean time $\LUV-\l$.  All solutions to the equation of motion $(\partial_y^2 +V_*)a=0$ derived from \eqref{Sa2d} can be written as
\beq
a(y)=c_1 \exp \left(\sqrt{-V_*}y\right) +c_2 \exp \left(-\sqrt{-V_*}y\right) \,.
\eeq
Fixing the constants $c_1$, $c_2$ by imposing the boundary conditions, and plugging the classical solution back to the action, we find the bare UV amplitude \eqref{PsiUV} to be
\beq\label{PUV2d}
\PUV^{(0)}=\mc N \exp\frac{\sqrt{-V_*}}{\kappa^2} \int d^2x \shg \,
\frac{(a_0^2+\ta^2) \cosh\left[ \sqrt{-V_*}(\LUV-\l) \right] -2a_0\ta}{\sinh\left[ \sqrt{-V_*}(\LUV-\l) \right]} \,,
\eeq
where $\mc N$ is a normalization constant that comes from both the determinant of the Gaussian integral and the last $\hR$ term in \eqref{Sa2d} (in cases where $\hR\neq 0$).  When we exponentiate $\mc N$, the part from the determinant is subdominant to $\WUV$ in the semiclassical limit $\kappa^2\to0$, and the part from $\hR$ is simply $\frac12 \hR (\LUV-\l)$ which does not depend on $\ta$.  Therefore we will neglect $\mc N$ in the following analysis.

It is easy to verify that \eqref{PUV2d} approaches the delta function $\delta(\ta-a_0)$ on the imaginary $a$-axis as $\l\to \LUV$, and also that it satisfies the HJ equation \eqref{wHJa} to be introduced in the next section (with $\phi$ treated as a background).

From the general expression \eqref{PUV2d} for $\PUV$, we may now wish to consider special cases.  For AdS cases ($V_*<0$), we may set $V_*=-1$ (and hence the AdS radius to 1) for simplicity.  We can also set the boundary condition as $a_0=\exp \LUV$ for flat slicing or $a_0=\sinh \LUV$ for dS slicing.  This corresponds to not introducing bare source terms for the trace $T$ of the stress tensor in the dual QFT.  Pushing the UV boundary $y=\LUV$ to the conformal boundary $y=\infty$ of AdS, we find that \eqref{PUV2d} simplifies to
\beq
\PUV^{(0)}=
\begin{cases}
\displaystyle \mc N \exp\frac{1}{\kappa^2} \int d^2x \shg \, (\ta^2-2 e^\l \ta+a_0^2)
\quad&\text{(AdS$_3$/dS$_2$)}\\
\displaystyle \mc N \exp\frac{1}{\kappa^2} \int d^2x \shg \, (\ta^2-4 e^\l \ta+a_0^2)
\quad&\text{(Poincar\'e AdS$_3$)}
\end{cases}
\eeq
Adding the appropriate counterterms, we have
\beq\label{PUVAdS3}
\PUV=
\begin{cases}
\displaystyle \mc N \exp\frac{1}{\kappa^2} \int d^2x \shg \, (2\ta^2-2 e^\l \ta)
\quad&\text{(AdS$_3$/dS$_2$)}\\
\displaystyle \mc N \exp\frac{1}{\kappa^2} \int d^2x \shg \, (2\ta^2-4 e^\l \ta)
\quad&\text{(Poincar\'e AdS$_3$)}
\end{cases}
\eeq

For the $dS_3/dS_2$ case ($V_*>0$), we may set $V_*=1$ and impose the boundary condition at the central slice $\LUV=\pi/2$.  The zero-source boundary condition is $a_0=\sin \LUV=1$.  Plugging these into \eqref{PUV2d}, we find that the UV amplitude simplifies to
\beq\label{PUVdS3}
\PUV=\mc N \exp\frac{1}{\kappa^2} \int d^2x \shg \, \frac{(\ta^2+1)\sin \l-2\ta}{\cos \l} \qquad \text{(dS$_3$/dS$_2$)}
\eeq
where we have used the fact that there are no counterterms in the dS case.

\subsection{Wilsonian action}

In this subsection, we calculate the Wilsonian action as a function of $T$ by directly evaluating the integral transform of \eqref{PUVAdS3} and \eqref{PUVdS3}:
\beq
\exp\left(-\kappa^{-2} s(T,\l)\right)= \int \cD \ta \exp\left\{ \frac{1}{\kappa^2} \int d^dx \shg \, \ba(\ta-\ba) T \right\} \PUV(\ta,\l) \,.
\eeq
In all cases the path integral is again Gaussian (with its wrong sign fixed by contour rotation).  As before we can neglect the determinant of the Gaussian integral because it does not contribute to the leading $\kappa^2\to0$ limit.

For $AdS_3$ with the zero-source boundary condition ($a_0=\sinh \LUV$ or $\exp \LUV$ as $\LUV\to\infty$, where we have rescaled $R_{(A)dS}$ to $1$), the Wilsonian action is
\beq\label{sigAdS3}
s=
\begin{cases}
\displaystyle \int d^2x \shg \, \left(\frac{\sinh^2\l}{8} T^2-\frac{1-e^{-2\l}}{4}T+\frac{1}{2}e^{2\l}\right)
\quad&\text{(AdS$_3$/dS$_2$)}\\
\displaystyle \int d^2x \shg \, \left(\frac{e^{2\l}}{8} T^2+2e^{2\l}\right)
\quad&\text{(Poincar\'e AdS$_3$)}
\end{cases}
\eeq

For the case of $dS_3/dS_2$ we impose the zero-source boundary condition ($a_0=\sin \LUV$) at the central slice $\LUV=\pi/2$, which leads to
\beq
s=\int d^2x \shg \, \left(\frac{\sin\l \cos\l}{4} T^2-(\cos^2 \l)T+\frac{\cos\l}{\sin\l}\right)
\qquad \text{(dS$_3$/dS$_2$)}
\eeq

These expressions can be straightforwardly generalized to arbitrary dimension, as long as we are working in the limit $\kappa^2 \rightarrow 0$.

\bibliographystyle{JHEP}

\begin{thebibliography}{10}

\bibitem{dSdS} 

 M.~Alishahiha, A.~Karch, E.~Silverstein and D.~Tong,
  ``The dS/dS correspondence,''
  AIP Conf.\ Proc.\  {\bf 743}, 393 (2005)
  [hep-th/0407125].

 M.~Alishahiha, A.~Karch and E.~Silverstein,
  ``Hologravity,''
  JHEP {\bf 0506}, 028 (2005)
  [hep-th/0504056].

\bibitem{dSdSbrane}

 X.~Dong, B.~Horn, E.~Silverstein and G.~Torroba,
  ``Micromanaging de Sitter holography,''
  Class.\ Quant.\ Grav.\  {\bf 27}, 245020 (2010)
  [arXiv:1005.5403 [hep-th]].

\bibitem{FRW} 
  X.~Dong, B.~Horn, S.~Matsuura, E.~Silverstein and G.~Torroba,
  ``FRW solutions and holography from uplifted AdS/CFT,''
  Phys.\ Rev.\ D {\bf 85}, 104035 (2012)
  [arXiv:1108.5732 [hep-th]].

\bibitem{AHM} 
  D.~Anninos, S.~A.~Hartnoll and D.~M.~Hofman,
  ``Static Patch Solipsism: Conformal Symmetry of the de Sitter Worldline,''
  Class.\ Quant.\ Grav.\  {\bf 29}, 075002 (2012)
  [arXiv:1109.4942 [hep-th]].

\bibitem{dSCFT} 
  D.~Anninos, T.~Hartman and A.~Strominger,
  ``Higher Spin Realization of the dS/CFT Correspondence,''
  arXiv:1108.5735 [hep-th].

  A.~Strominger,
  ``Inflation and the dS / CFT correspondence,''
  JHEP {\bf 0111}, 049 (2001)
  [hep-th/0110087].

  A.~Strominger,
  ``The dS / CFT correspondence,''
  JHEP {\bf 0110}, 034 (2001)
  [hep-th/0106113].


\bibitem{HarlowSusskind} 
  D.~Harlow and L.~Susskind,
  ``Crunches, Hats, and a Conjecture,''
  arXiv:1012.5302 [hep-th].



\bibitem{HP} 
  I.~Heemskerk and J.~Polchinski,
  ``Holographic and Wilsonian Renormalization Groups,''
  JHEP {\bf 1106}, 031 (2011)
  [arXiv:1010.1264 [hep-th]].


\bibitem{Mansfield:1999kk} 
  P.~Mansfield and D.~Nolland,
  ``One loop conformal anomalies from AdS / CFT in the Schrodinger representation,''
  JHEP {\bf 9907}, 028 (1999)
  [hep-th/9906054].


\bibitem{de Boer:1999xf} 
  J.~de Boer, E.~P.~Verlinde and H.~L.~Verlinde,
  ``On the holographic renormalization group,''
  JHEP {\bf 0008}, 003 (2000)
  [hep-th/9912012].

\bibitem{Li:2000ec} 
  M.~Li,
  ``A Note on relation between holographic RG equation and Polchinski's RG equation,''
  Nucl.\ Phys.\ B {\bf 579}, 525 (2000)
  [hep-th/0001193].

\bibitem{Akhmedov:2010sw} 
  E.~T.~Akhmedov and E.~T.~Musaev,
  ``An exact result for Wilsonian and Holographic renormalization group,''
  Phys.\ Rev.\ D {\bf 81}, 085010 (2010)
  [arXiv:1001.4067 [hep-th]].
  
\bibitem{Faulkner:2010jy} 
  T.~Faulkner, H.~Liu and M.~Rangamani,
  ``Integrating out geometry: Holographic Wilsonian RG and the membrane paradigm,''
  JHEP {\bf 1108}, 051 (2011)
  [arXiv:1010.4036 [hep-th]].

\bibitem{inprogress}
M. Dodelson et al., work in progress.  


\bibitem{FRWCFT} 
 B.~Freivogel, Y.~Sekino, L.~Susskind and C.~-P.~Yeh,
 ``A Holographic framework for eternal inflation,''
 Phys.\ Rev.\ D {\bf 74}, 086003 (2006)
 [hep-th/0606204].

\bibitem{functionalRG}
  K.~G.~Wilson,
  ``The renormalization group and critical phenomena,''
  Rev.\ Mod.\ Phys.\  {\bf 55}, 583 (1983).

J.~Polchinski,
  ``Renormalization and Effective Lagrangians,''
  Nucl.\ Phys.\ B {\bf 231}, 269 (1984);

C.~Wetterich,
  ``Exact evolution equation for the effective potential,''
  Phys.\ Lett.\ B {\bf 301}, 90 (1993).

\bibitem{Trefs}

 M.~Henningson and K.~Skenderis,
  ``The Holographic Weyl anomaly,''
  JHEP {\bf 9807}, 023 (1998)
  [hep-th/9806087].

\bibitem{HS}
 D.~Harlow and D.~Stanford,
  ``Operator Dictionaries and Wave Functions in AdS/CFT and dS/CFT,''
  arXiv:1104.2621 [hep-th].

\bibitem{orbifold}

 S.~Kachru and E.~Silverstein,
  ``4-D conformal theories and strings on orbifolds,''
  Phys.\ Rev.\ Lett.\  {\bf 80}, 4855 (1998)
  [hep-th/9802183];

A.~Dymarsky, I.~R.~Klebanov and R.~Roiban,
  ``Perturbative search for fixed lines in large N gauge theories,''
  JHEP {\bf 0508}, 011 (2005)
  [hep-th/0505099].

S.~Kachru, D.~Simic and S.~P.~Trivedi,
  ``Stable Non-Supersymmetric Throats in String Theory,''
  JHEP {\bf 1005}, 067 (2010)
  [arXiv:0905.2970 [hep-th]].

\bibitem{bulkobs}

A.~Hamilton, D.~N.~Kabat, G.~Lifschytz and D.~A.~Lowe,
  ``Local bulk operators in AdS/CFT: A Boundary view of horizons and locality,''
  Phys.\ Rev.\ D {\bf 73}, 086003 (2006)
  [hep-th/0506118].

I.~Heemskerk, D.~Marolf and J.~Polchinski,
  ``Bulk and Transhorizon Measurements in AdS/CFT,''
  arXiv:1201.3664 [hep-th].

\bibitem{BF} 
  P.~Breitenlohner and D.~Z.~Freedman,
  ``Positive Energy in anti-De Sitter Backgrounds and Gauged Extended Supergravity,''
  Phys.\ Lett.\ B {\bf 115}, 197 (1982).

  P.~Breitenlohner and D.~Z.~Freedman,
  ``Stability in Gauged Extended Supergravity,''
  Annals Phys.\  {\bf 144}, 249 (1982).

\bibitem{Coleman}
	S.~Coleman,
	``Aspects of Symmetry: Selected Erice Lectures,''
	Cambridge University Press, 420 pages, 1988.


\bibitem{Anninos:2011zn} 
  D.~Anninos, T.~Anous, I.~Bredberg and G.~S.~Ng,
  ``Incompressible Fluids of the de Sitter Horizon and Beyond,''
  JHEP {\bf 1205}, 107 (2012)
  [arXiv:1110.3792 [hep-th]].
  
\bibitem{Anninos:2012qw} 
  D.~Anninos,
  ``De Sitter Musings,''
  Int.\ J.\ Mod.\ Phys.\ A {\bf 27}, 1230013 (2012)
  [arXiv:1205.3855 [hep-th]].
  
\bibitem{Bousso:2009mw} 
  R.~Bousso and I-S.~Yang,
  ``Global-Local Duality in Eternal Inflation,''
  Phys.\ Rev.\ D {\bf 80}, 124024 (2009)
  [arXiv:0904.2386 [hep-th]].

D.~Harlow, S.~H.~Shenker, D.~Stanford and L.~Susskind,
  ``Tree-like structure of eternal inflation: A solvable model,''
  Phys.\ Rev.\ D {\bf 85}, 063516 (2012)
  [arXiv:1110.0496 [hep-th]].


\end{thebibliography}
\renewcommand{\refname}{Bibliography}
\addcontentsline{toc}{section}{Bibliography}
\providecommand{\href}[2]{#2}\begingroup\raggedright

\end{document}